\RequirePackage{fix-cm}

\documentclass{svjour3}                     

\smartqed  

\usepackage{graphicx}

\usepackage{natbib}

\usepackage[hyphens]{url}

\newcommand{\subparagraph}{}

\usepackage{titlesec}

\journalname{Scientometrics}

\begin{document}

\title{Prediction of Emerging Technologies Based on Analysis of the U.S.
Patent Citation Network
\thanks{PE thanks  the Henry Luce Foundation and the Toyota Research Institue for their support. KJS acknowledges the generous support of The Filomen D'Agostino and Max E. Greenberg Research Fund.}
}

\titlerunning{Prediction of Emerging Technologies}        

\author{P\'eter \'Erdi 		\and
	Kinga Makovi		\and
	Zolt\'an Somogyv\'ari	\and 
	Katherine Strandburg	\and
	Jan Tobochnik		\and
	P\'eter Volf		\and
	L\'aszl\'o Zal\'anyi
}


\institute{P. \'Erdi \and K. Makovi \and J. Tobochnik \and L. Zal\'anyi \at
              Center for Complex Systems Studies, Kalamazoo College, Kalamazoo, MI 49006, USA \\
              Tel.: +(269) 337-5720\\
              Fax: +(269) 337-7101\\
              \email{perdi@kzoo.edu}         
           \and
           P. \'Erdi \and K. Makovi \and Z. Somogyv\'ari \and P. Volf \and L. Zal\'anyi \at
              Complex Systems and Computational Neuroscience Group, Wigner Research Centre for Physics, Hungarian Academy of Sciences, Budapest, Hungary
	   \and
	   P. Volf \at
              Department of Measurement and Information Systems, Budapest University of Technology and Economics, Budapest, Hungary \\
	      Network and Subscriber Data Management, Nokia Siemens Networks, Budapest, Hungary
	   \and
	   K. Makovi \at
	      Department of Sociology, Columbia University, New York, NY, USA
	   \and
	   K. Strandburg \at
	      New York University School of Law, New York, NY, USA
}

\date{Received: date / Accepted: date}

\maketitle

\noindent Correspondence to P\'eter \'Erdi, address: Kalamazoo College,
1200 Academy Street, Kalamazoo, MI 49006, USA; phone: +(269) 337-5720;
fax:
+(269) 337-7101; \mbox{e-mail:} perdi@kzoo.edu

\begin{abstract}
\noindent
The network of patents connected by citations is an evolving graph, which
provides a representation of  the innovation process. A patent citing another
implies that the cited patent reflects a piece of previously existing
knowledge that the citing patent builds upon. A methodology presented here
(i) identifies  actual  clusters of patents: i.e. technological branches,
and  (ii) gives predictions about the temporal changes of the structure of
the clusters. A predictor, called the {citation vector}, is defined
for characterizing technological development to show how a patent cited by
other patents belongs to various industrial fields. The clustering
technique adopted is able to detect the new emerging recombinations, and
predicts emerging new technology clusters. The predictive ability
of our new method is illustrated on the example of USPTO subcategory 11,
Agriculture, Food, Textiles. A cluster of patents is determined based
on citation data up to 1991, which shows significant overlap of the class
442 formed at the beginning of 1997. These new tools of predictive
analytics could support policy decision making processes in science and
technology, and help formulate recommendations for action.

\keywords{patent citation \and network \and co-citation clustering \and technological evolution}
\end{abstract}

\section{Introduction}

In this paper we present a conceptual framework and a computational algorithm for studying the process of technological evolution and making predictions about it by mining the patent citation network. Patent data long has been recognized as a rich and potentially fruitful source of information about innovation and technological change. Besides describing and claiming inventions, patents cite previous patents (and other references) that are relevant to determining whether the invention is sufficiently novel and nonobvious to be patented.  Citations are contributed by patentees, patent attorneys, and patent office examiners.  Patents, as nodes, and citations between them, as edges, form a growing directed network, which aggregates information about technological relationships and progress provided by those players. Our methodology seeks to detect incipient technological trends reflected in the citation network and thus to predict their emergence. The proposed method should also be useful for analyzing the historical evolution of patented technology.

Innovation is frequently deemed key to economic growth and sustainability \citep{saviotti03,saviotti05b,saviotti05}. Often, economically significant technologies are the result of considerable investment in basic research and technology development. Basic research is funded by the government and, to a lesser extent, by large firms such as those found in the medical and pharmaceutical industries. Development is carried out by a range of players, including start-up companies spending large fractions of their revenues on innovation. Not surprisingly research and development (R\&D) costs have risen rapidly in the past few decades. For example, ``worldwide R\&D expenditures in 2007 totaled an estimated \$1,107 billion.''\footnote{http://www.nsf.gov/statistics/seind10/c4/c4s5.htm}

Because innovation is unpredictable, R\&D investment is often risky. The practical implications for particular firms are evident:  ``The continuous emergence of new technologies and the steady growth of most technologies suggest that relying on the status quo is deadly for any firm...'' \citep{sood-tellis05}. As Day and Schoemaker argue \citep{day-smaker05}: ``...The biggest dangers to a company are the ones you don't see coming. Understanding these threats -- and anticipating opportunities -- requires strong peripheral vision.'' In the long term, understanding the emergence of new technological fields could help to orient public policy, direct investment, and reduce risk, resulting in improved economic efficiency.  Detecting the emergence of new technological branches is an intrinsically difficult problem, however.

Recent improvements in computing power and in the digitization of patent data make possible the type of large-scale data mining methodology we develop in this paper. Our approach belongs to the field of predictive analytics, which is a branch of data mining concerned with the prediction of future trends. The central element of predictive analytics is the predictor, a mathematical object that can be defined for an individual, organization or other entity and employed to predict its future behavior. Here we define a `citation vector' for each patent to play the role of a predictor. Each coordinate  of the citation vector is proportional to the relative frequency that the patent has been cited by other patents in a particular technological category at a specific time. Changes in this citation vector over time reflect the changing role that a particular patented technology is playing as a contributor to later technological development.  We hypothesize that patents with similar citation vectors will belong to the same technological field.  To track the development and emergence of technological clusters, we employ clustering algorithms based on a measure of similarity defined using the citation vectors. We identify the community structure of non-assortative patents -- those which receive citations from outside their own technological areas. The formation of new clusters should correspond to the emergence of new technological directions.

A predictive methodology should be able to ``predict'' evolution from the ``more distant'' past to the ``more recent'' past. A process, called backtesting checks if this criteria is met, which if true, holds the promise of predictions from the present to the near future. To illustrate the potential of our approach and demonstrate that the emergence of new technological fields can be predicted from patent citation data, we backtested by using our method to ``predict'' an emerging technological area that was later recognized as a new technological class by the US Patent and Trademark Office (USPTO).

Because the patent citation network reflects social activity, the potential scope and limitations of prediction are different from those in the natural sciences. Unexpected scientific discoveries, patent laws, habits of patent examiners, the pace of economic growth and many other factors influence the development of technology and of the patent network that we do not intend to explicitly model. Correspondingly, patent grants change the innovative environment, which we also don't incorporate in our modeling strategy. Any predictive method can peer only into the relatively short term future, ours is not an exception. The methodology developed here harnesses assessments of technological relationships made by a very large number of participants in the system and attempts to capture the larger picture emerging from those grass root assessments. We hope to show what is possible based on a purely structural analysis of the patent citation data in spite of the above mentioned difficulties.

\section{The Patent Citation Network}

The United States patent system is a very large compendium of information about technology, and its evolution goes back more than two hundred years.  It contains more than 8 million patents.  The system reflects technological developments worldwide.  Currently about half of US patents are granted to foreign inventors.  Of course, the US patent system is not a complete record of technological evolution:  not all technological developments are eligible for patenting and not all eligible advances are patented in the United States or anywhere else, for that matter.  Nonetheless, the United States patent system is a well-studied and documented source of data about the evolution of technology.  Thus, we have chosen it as the primary basis of our investigation, keeping in mind that our method potentially could be applied to other patent databases as well.

Complex networks have garnered much attention in the last decade. The application of complex network analysis to innovation networks has provided a new perspective from which to understand the innovation landscape \citep{pyka09}.  In our study, the patent citation network is comprised of patents (nodes) and the citations between them (links). A patent citation reflects a technological relationship between the inventions claimed in the citing and cited patents.  Citations are contributed by patentees and their attorneys and by patent examiners.  They reflect references to be considered in determining whether the claimed invention meets the patentability requirements of novelty and nonobviousness.  Both patentees and patent examiners have incentives to cite materially related prior patents. Patent applicants are legally required to list related patents of which they are aware. Patent examiners seek out the most closely related prior patents so that they can evaluate whether a patent should be granted. Consequently, citation of one patent by another represents a technological connection between them, and the patent citation network reflects information about technological connections known to patentees and patent examiners.  Patents sometimes cite scientific journals and other non-patent sources.  We ignore those citations here.  Taking them into account is not necessary to our goal of identifying emerging technology clusters and is not possible using our current methodology, though it may be possible in the future to improve the method by devising a means to take them into account. In the literature review section we will give a more detailed analysis of what one can infer from the patent network, here we cite only \citep{duguet05}:''...patent citations are indeed related to firms' statements about their acquisition and dispersion of new technology...''.

As described below, our methodology utilizes a classification system based on the one that USPTO uses in defining our citation vector. However, our methodology has the promise to predict the emergence of new technological fields not yet captured by the USPTO classification system. The USPTO system has about 450 classes, and over 120,000 subclasses. All patents and published patent applications are manually assigned to primary and secondary classes and subclasses by patent examiners.  The classification system is used by patent examiners and by applicants and their attorneys and agents as a primary resource for assisting them in searching for relevant prior art. Classes and sub-classes are subject to ongoing modification, reflecting the USPTO's assessment of technological change. Not only are new classes added to the system, but patents can be reclassified when a new class is defined. As we discuss later, that reclassification provides us with a natural experiment, which offers an opportunity to test our methodology for detecting emerging new fields \citep{jaffe-trajt05}. Within the framework of a project sponsored by the National Bureau of Economic Research (NBER), a higher-level classification system was developed, in which the 400+ USPTO classes were aggregated into 36 subcategories\footnote{11 -- Agriculture, Food, Textiles; 12 -- Coating; 13 -- Gas; 14 -- Organic Compounds; 15 -- Resins; 19 -- Miscellaneous-Chemical; 21 -- Communications; 22 -- Computer Hardware\&Software; 23 -- Computer Peripherials; 24 -- Information Storage; 31 -- Drugs; 32 -- Surgery\&Med Inst; 33 -- Biotechnology; 39 -- Miscellaneous-Drgs\&Med; 41 -- Electrical Devices; 42 -- Electrical Lighting; 43 -- Measuring\&Testing; 44 -- Nuclear\&X-rays; 45 -- Power Systems; 46 --Semiconductor Devices; 49 -- Miscellaneous-Electric; 51 -- Mat.Proc\&Handling; 52 -- Metal Working; 53 -- Motors\&Engines+Parts; 54 -- Optics; 55 -- Transportation; 59 -- Miscellaneous-Mechanical; 61 -- Agriculture, Husbandry, Food; 62 -- Amusement Devices; 63 -- Apparel\&Textile; 64 -- Earth Working\&Wells; 65 -- Furniture, House, Fixtures; 66 -- Heating; 67 -- Pipes\&Joints, 68 -- Receptacles, 69 -- Miscellaneous-Others.}, which were further lumped into six categories (Computers and Communications, Drugs and Medical, Electrical and Electronics, Chemical, Mechanical and Others).  As any classification system, this system reflects ad hoc decisions as to what constitutes a category or a subcategory. However this classification system appears to show sufficient robustness to be of use in our methodology \citep{jaffe-trajt05}.

\section{Literature Review}

Two strands of literature form the context for our research.  First, there is a large literature in which patent citations (and, relatedly, academic journal citations) are used to explore various aspects of technological and scientific development.  Second, there is a literature involving attempts to produce predictive roadmaps of the direction of science and technology.  Our project, along with a few others, stands at the intersection of these two literatures, using patent citations as a predictive tool.

\subsection{Patent citation analysis}

Patent citation analysis has been used for a variety of different objectives.  For example, citation counts have long been used to evaluate research performance \citep{garfield83,moed05}. Economic studies suggest that patent citation counts are correlated with economic value \citep{harhoff99,sampat-ziedonis02,hagedoorn03,lanjouw-schankerman04,jaffe-trajt05}.  More generally, patent citation data has been used in conjunction with other empirical information, such as information about the companies who own patents, interviews with scientists in the field, and analysis of the citation structure of scientific papers, to explore the relationship between innovation and the patent system \citep{narin94,milman94,meyer01,kostoff-schaller01,debackere02,murray02,verbeek02}.  Patent citations have also been used to investigate knowledge flows and spillovers \citep{duguet05,strumsky05,fleming06,sorenson06}, though the validity of these studies is called into question by the fact that patent citations are very frequently inserted as a result of a search for relevant prior art by a patentee's attorney or agent or a patent examiner \citep{sampat04,criscuolo08,alcacer06}. From previous work the basic idea is that new knowledge comes from combinations of old knowledge. Local combinations are more effective than distant combinations, which though rare, when they occur provide major new knowledge \citep{sternitzke09}. We ourselves have used the methodologies of modern network science to study the dynamic growth of the patent citation network in an attempt to better understand the patent system itself and the possible effects of changes in legal doctrine \citep{strandburg07,strandburg09}.

Other scholars have used patents as proxies for invention in attempts to develop and test theoretical models for the process of technological evolution. For example, Fleming and Sorenson \citep{fleming01,fleming01b} develop a model of technological evolution in which technology evolves primarily by a process of searching for new combinations of existing technologies. In analogy to biological evolution, they imagine a ``fitness landscape'' and a process of ``recombinant search'' by which technological evolution occurs. They use citations as a measure of fitness, and conclude that a successful innovation balances between re-using familiar components -- an approach which is likely to succeed -- and combining elements that have rarely been used together -- an approach that often fails, but produces more radical improvements. In a later paper Fleming \citep{fleming04} uses patent citation data to explore the value of using science to guide technological innovation by tracking the number of patent citations to non-patent sources, and measures the difficulty of an invention by looking at how subclasses related to the patent were previously combined. His main conclusion is that science provides the greatest advantage to those inventions with the greatest coupling between the components (non-modularity). Podolny and Stuart also use patent citations to study the innovative process \citep{podolny95}. They define local measures of competitive intensity and competitive crowding in a technological niche based on indirect patent citation ties and use these measures to study the ways in which a technological niche can become crowded and then exhausted as innovative activity proceeds.

Most relevant to our work here is that patent citations, as well as academic journal citations, have been used to study the ``structure'' of knowledge as reflected in different fields and sub-fields.  All of these methods rely on some way to measure the similarity or relatedness of patents or journal articles.  Co-citation analysis, one approach to this problem, goes back to the now classic works of Small \citep{small73,garfield93}:  ``...A new form of document coupling called co-citation is defined as the frequency with which two documents are cited together. The co-citation frequency of two scientific papers can be determined by comparing lists of citing documents in the Science Citation Index and counting identical entries. Networks of co-cited papers can be generated for specific scientific specialties [...] Clusters of co-cited papers provide a new way to study the specialty structure of science...''  The assumption of co-citation analysis is that documents that are frequently cited together cover closely related subject matter.  Co-citation analysis has been used recently \citep{chenetal10}, as well as more than a decade ago \citep{mogee-kolar98a,mogee-kolar98b,mogee-kolar99} to study the structure of knowledge in various specific fields, such as nanotechnology \citep{huang03,huangetal04,meyer01,kostoff06}, semiconductors \citep{almeida-kogut97}, biotechnology \citep{mcmillan00}, and tissue engineering \citep{murray02} providing valuable insight into the development of these technological fields. The main goal of this line of research is to understand in detail the development of a specific industrial sector.  Wallace et al.\ \citep{wallaceetal09} recently adopted a method \citep{blondeletla08} for using co-citation networks to detect clusters that relies on the topology of the citation-weighted network. Lai and Wu \citep{laiwu05} have argued that co-citation might be used to develop a patent classification system to assist patent managers in understanding the basic patents for a specific industry, the relationships among categories of technologies, and the evolution of a technology category -- arguing very much along the lines that we do here.

Researchers are also exploring the use of concepts taken from modern network science and social network studies to illuminate the structure of technology using the patent citation network.  For example, Weng et al.\ \citep{weng10}, employ the concept of structural equivalence, a fundamental notion in the classical theory of social networks. They define two patents as ``structurally equivalent'' in the technological network when they cite the same preceding patents.  They use structural equivalence to map out the relationships between 48 insurance business method patents and classify patents as belonging to the technological core or periphery.  They compare their results to a classification made by expert inspection of the patents. Our predictor, the `citation vector' described in the following section can be conceived as a tool to get at a weighted version of structural equivalence.

Other researchers have measured the distance between patents as defined by the shortest path along the patent citation network.  For example, Lee at al.\ \citep{lee10} recently analyzed a small subset of the patent citation network  to study the case of electrical conducting polymer nanocomposites. Chang et al.\ \citep{chang09} use a measure of ``strength'' based on the frequency of both direct and indirect citation links to define a small set of ``basic'' patents in the business method arena and then use clustering methods to determine the structure of relationships among those basic patents.  For the most part, these studies have been limited to small numbers of patents and many have focused primarily on visualization.

Researchers have also used citations between patents and the scientific literature to investigate the relationship between scientific research and patented technology.  For example, a comparative study \citep{shibata10} of the structures of the scientific publication and patent citation networks in the field of solar cell technology found a time lag between scientific discovery and its technological application.  Other researchers have also considered the role of science in technological innovation by investigating citations between patents and articles in the scientific literature \citep{mogee-kolar98a,mcmillan00,meyer01,tijssen01,fleming04}. One main difference between scientific and patent citations is that they are less likely to have identical references. To account for this difference, it was suggested that the patentee and the patent examiner reference more objectively prior art that is relevant, whereas authors of journal articles have motivation to reference papers that are irrelevant to the subject of the study \citep{meyer00}.

Finally, there are various non-citation-based methods of determining the similarity between patents (or between academic articles).  Researchers have used text mining \citep{huang03,kostoff06}, keyword analysis \citep{huangetal04}, and co-classification analysis based on USPTO classifications \citep{leydesdorff08} in this way. These approaches are usually more time consuming, their implementation may be specific to a technological field and involve ad hoc decisions in the classification process and thus they are less systematic than ours. Generally, while we don't believe that patent citation is \emph{the} magic bullet to identify the most promising emerging technologies, we see that the more traditional methods of patent trajectory analysis and the new methods are converging \citep{fontana09}.

\subsection{Predicting the Direction of Science and Technology}

There is a large literature concerning various approaches to producing ``roadmaps'' of science and technology as tools for policy and management based on numerous approaches ranging from citation analysis to expert opinion  (see, e.g., \citet{kostoff-geisler07,kajikawa08a}, and references therein).  Some of this work seeks to provide direction to specific industrial research and development processes.  Recently, for example, OuYang and Weng \citep{ouyang11} suggested the use of patent citation information, in addition to other information such as expert analysis, during the process of new product design.  Citation-based methods, specifically co-citation clustering techniques, have been employed for tracking and predicting growth areas in science using the scientific literature \citep{small06}. Because the patent system is such a rich source of information about the evolution of technology,
it has long been hoped that insights into the mechanisms of technological changes can be  used to make predictions on emerging fields of technologies \citep{breitzman07}.

The extensive work of Kajikawa and his coworkers, which focuses on mining larger scale trends from citation data, is most closely related to the work we report here. These studies use citations to cluster scientific articles in order to detect emerging research areas.  These researchers have deployed various clustering techniques based on co-citation and on direct citation networks \citep{shibata08}, to explore research evolution in the sustainable energy industry (fuel cell, solar cell) \citep{kajikawa08a}, the area of biomass and bio-fuels \citep{kajikawa08c}, and, most recently, in the field of regenerative medicine \citep{shibata11}.

\section{Research methodology}

\subsection{Evolving clusters}

\begin{figure}
\begin{centering}
\includegraphics[width=0.65\textwidth]{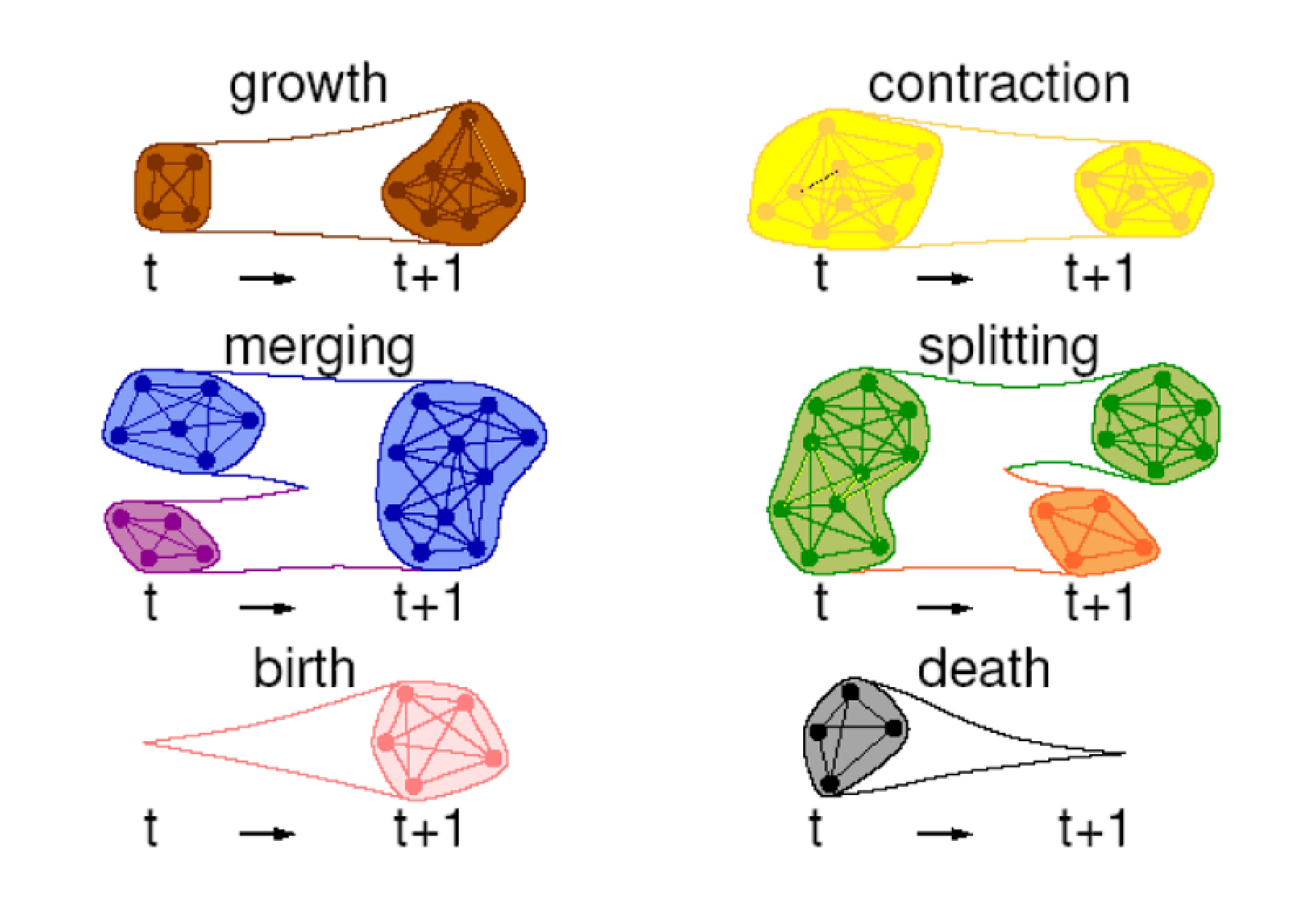}
\caption{\textbf{Possible elementary events of cluster evolution}. Based on
Palla et al.\ \citep{palla07}.
\label{fig:palla}}
\end{centering}
\end{figure}

The basic orientation of our research is similar to that of Kaijikawa and co-coworkers.  We search for emerging and evolving technology clusters based on a citation network.  Our method differs from previous work in several respects, however.  All citation-based clustering methods leverage the grass roots, field-specific expertise embedded in the citation network.  Here, because we are clustering patents we are able to make use of the additional embedded expertise reflected in the assignment of patents to USPTO classifications by patent examiners.  In our clustering approach, we use evolving patterns of citations to a particular patent by other patents in various technology categories to measure patent similarity.  Our method can be used to analyze large subsets of the patent citation network in a systematic fashion and to observe the dynamics of cluster formation and disappearance over time.  In the long run we hope to be able to describe these dynamics systematically in terms of birth, death, growth, shrinking, splitting and merging of clusters, analogous to the cluster dynamical elementary events described by Palla et al.\ \citep{palla07}. Figure\,\ref{fig:palla} illustrates these events.

Although our method is based on USPTO's and NBER's classification of patents, we believe, that any classification system, which covers the whole technological space and describes it in sufficient detail could serve as the basis for our investigation.

\subsection{Construction of a predictor for technological development}

To capture the time evolution of technological fields, we have constructed a quantity, the `citation vector', which we use to define measures of similarity between patents.

Specifically, we define the citation vector for a given patent at any given time in the following way:

\begin{enumerate}
	\item For each patent, we calculate the sum of the citations received by that patent from patents in each of the 36 technological subcategories defined by Hall et al.\ \citep{hall01}; as noted above, these subcategories are aggregations of USPTO patent classifications. This gives us 36 sums for each patent, which we treat as entries in a 36-component vector. In the process of calculating the sums we weight incoming citations with respect to the overall number of citations made by the sender patent, thus, we give more weight to senders which referred to fewer others. The coordinate corresponding to each patent's own subcategory is set to zero so that the citation vectors focus on the combination of different technological fields.
	\item We then normalize the 36-component vector obtained for each patent in the previous step using an Euclidean norm to obtain our citation vector. Patents that have not received any citations are assigned a vector with all zero entries.  The citation vector's components may be interpreted as describing the relative influence that a patent has had on different technological areas at a specific time.  The impact of a patent on future technologies changes over time, and thus the citation vector evolves to reflect the changing ways in which a patented invention is reflected in different technological fields.
\end{enumerate}

We next seek to group patents into clusters based on their roles in the space of technologies. To do this, we define the similarity between two patents as the Euclidean distance of their citation vectors and apply clustering algorithms based on this similarity measure.  We thus hypothesize that patents cited in the same proportion by patents in other technological areas have similar technological roles.  This hypothesis resonates with the hypothesis formulated by the pioneer of co-citation analysis, Henry Small \citep{small73}, that co-cited patents are technologically similar.

Our focus is on those inventions that were influential in technologies other than their own. In other words, we are concentrating on patents that received non-assortative citations \citep{newman02}.  Because the high number of patents receiving only intra-subcategory citations tends to mask the recombinant process, citations within the same subcategory are eliminated from the citation vector.

Our algorithm for predicting technological development consists of
the following steps:

\begin{enumerate}
	\item Select  a time point $t_{1}$ between 1975 and 2007 and drop all patents that were issued after $t_{1}$.
	\item Keep some subset of subcategories: $c_{1}, c_{2}, \dots, c_{n}$ -- to work with a reasonably sized problem.
	\item Compute  the citation vector. Drop patents with assortative citation only.
	\item Compute the similarity matrix of patents by using the Euclidean distance product between the corresponding citation vectors.
	\item Apply a hierarchical clustering algorithm to reveal the functional clusters of patents.
	\item Repeat the above steps for several time points $t_{1} < t_{2} < \dots < t_{n}$.
	\item Compare the dendrogram obtained by the clustering algorithm for different time points to identify structural changes (such as emergence and/or disappearance of groups).
\end{enumerate}

The discussion thus far leaves us with two key questions: (i) What algorithms should be chosen to cluster the patents? (ii) How should we link the clustering results from consecutive time steps? The following subsections address these questions.

\begin{table*}
\caption{Number of patents in the examined networks and subnetworks in different moments. 
\label{fig:tab1}}
\begin{tabular}{ l | c c c c}
						&01.01.91	& 01.01.94	& 01.01.97	& 12.31.99 		\\
\hline\noalign{\smallskip}
Patents in the whole database 			&4980927	& 5274846 	& 5590420 	& 6009554	\\
Patents in the subcategory 11 			&18833		& 21052 	& 23191 	& 25624 	\\
Patents belong to class 442 from 1997 		&2815		& 3245 		& 3752 		& 4370 		\\
Patents with non-zero citation vector in 11	& 7671 		& 9382 		& 11245 	& 13217		\\
citations connected to patents in SC 11		& 70920		& 92177 	& 120380 	& 161711  	\\
\end{tabular}	
\end{table*}

\subsection{Identification of patent clusters}

Several clustering and graph partitioning algorithms are reasonable candidates for our project. An important pragmatic constraint in choosing clustering algorithms is their computation-time complexity. Given the fact that we are working on a huge database, we face an unavoidable trade-off between accuracy and computation time. Because we do not know a priori the appropriate number of clusters, hierarchical methods are preferred, as they do not require that the number of clusters be specified in advance. Available clustering methods include the k-means and Ward methods, which are point clustering algorithms \citep{ward63}. Graph clustering algorithms, such as those that use edge-betweenness \citep{girvan02,newman04} random walks \citep{pons06} and the MCL method \citep{dongen00} are also possible choices. The otherwise celebrated clique-percolation method \citep{palla07} employs a very restrictive concept of a k-clique, making it difficult to mine clusters from the patent citation network. Spectral methods \citep{newman06} are not satisfactory because they are extremely computation-time intensive in cases like this one, in which one would have to calculate the eigenvalues of a relatively dense  matrix. In the application presented here we adopted the Ward method.

\subsection{Detecting structural changes in the patent cluster system}

The structure of dendrograms resulting from hierarchical clustering methods, such as the Ward method, reflects structural relationships between patent clusters. In this hierarchy, each branching point is binary and defined only by its height on the dendrogram, corresponding to the distance between the two branches. Thus, all types of temporal changes in the cluster structure can be divided into four elementary events: 1) increase or 2) decrease in the height of an existing branching point, and 3) insertion of a new or  4) fusion of two existing branching points. To find these substantial, structural changes, we identify the corresponding branching points in the dendrograms representing consecutive time samples of the network and follow their evolution through the time period documented in the database.

To test whether our clusters are meaningful, we can compare the emergence of new clusters to the introduction of new classes by the USPTO. Potential new classes can be identified in the clustering results by comparing the dendrogram structure with the USPTO classification. While some of the branching points of the dendrogram are reflected in the current classification structure, we may find significant branches which are not identified by the classification system used at that time point and test our approach by seeing whether clusters that emerge at a particular time are later identified as new classes by the USPTO.

\section{Results and model validation}

\begin{figure*}
\begin{centering}
\includegraphics[width=0.75\textwidth]{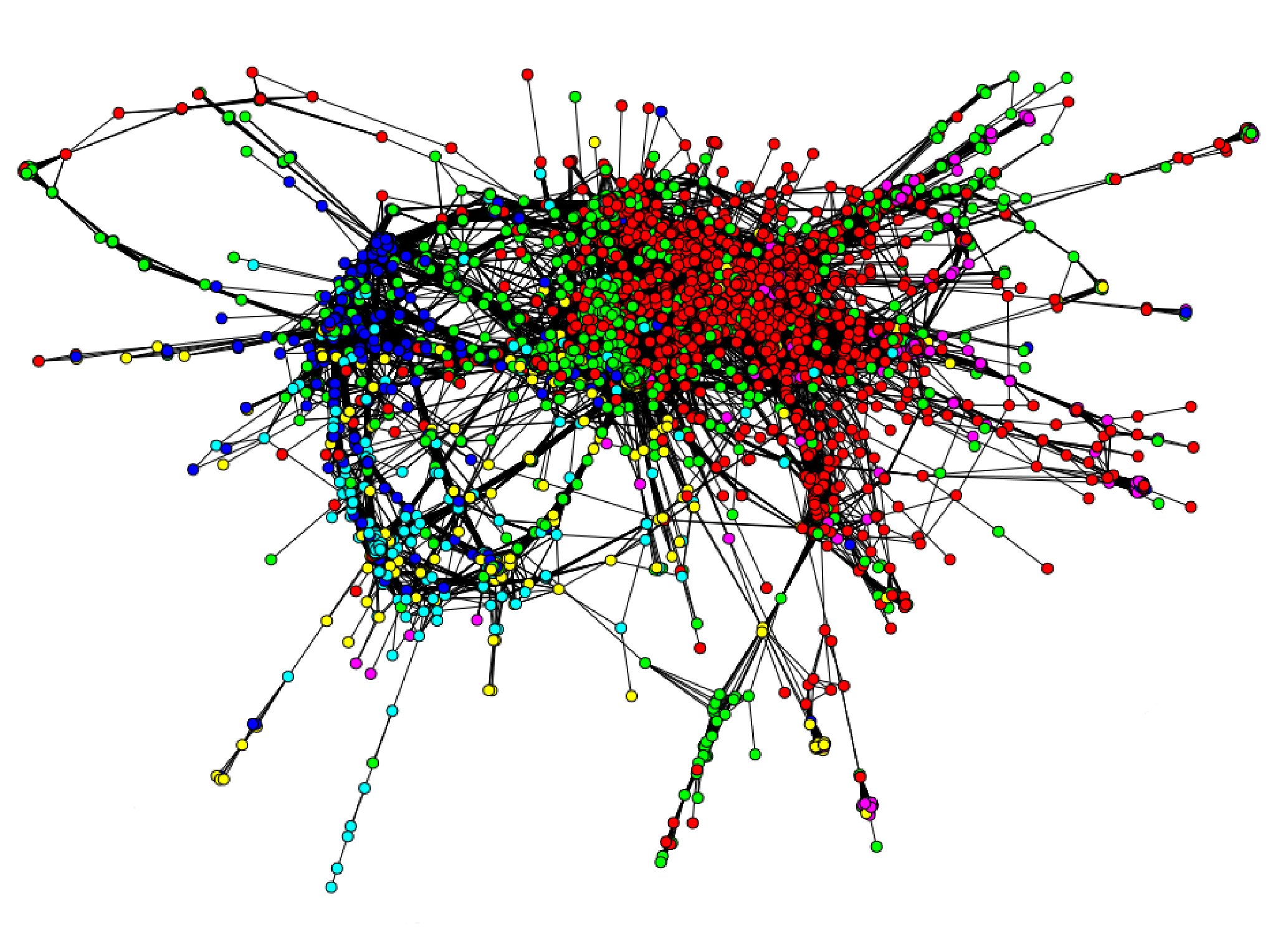}
\caption{\textbf{Cluster structure of patents in the citation space.}
Two-dimensional representation of patent similarity structure in the
subcategory 11 by using the Fruchterman-Reingold algorithm. Local
densities corresponding to technological areas can be recognized by the naked
eye or identified by clustering methods. The colors encode the US patent
classes: red corresponds to class 8; green: 19; blue: 71; magenta: 127;
yellow: 442; cyan: 504.
\label{fig:cluster}}
\end{centering}
\end{figure*}

We have chosen the NBER subcategory 11, Agriculture, Food, Textiles as an example, to demonstrate our method. The rationals of our choice are:

\begin{enumerate}
	\item Subcategory 11 (SC 11) has moderate size (compared to other subcategories), which was appropriate to the first test of our algorithm.
	\item SC 11 is heterogeneous enough to show non-trivial structure.
	\item A new USPTO class, the class 442 was established recently within the	subcategory 11, which we can use to test our approach.
\end{enumerate}

Note, that restriction of the field of investigation does not restrict the possibility of cross-technological interactions, because the citation vector remains 36 dimensional, including all the possible interactions between the actually investigated and all the other technological fields.

\subsection{Patent clusters: existence and detectability}

We begin by demonstrating the existence of local patent clusters based on the citation vector. Such clusters can be seen even with the naked eye by perusing a visualization of the 36 dimensional citation vector space projected onto two dimensions, or can be extracted by a clustering algorithm. See Figure\,\ref{fig:cluster}.

\subsection{Changes in the structure of clusters reflects technological evolution}

\begin{figure*}
\begin{centering}
\includegraphics[width=0.85\textwidth]{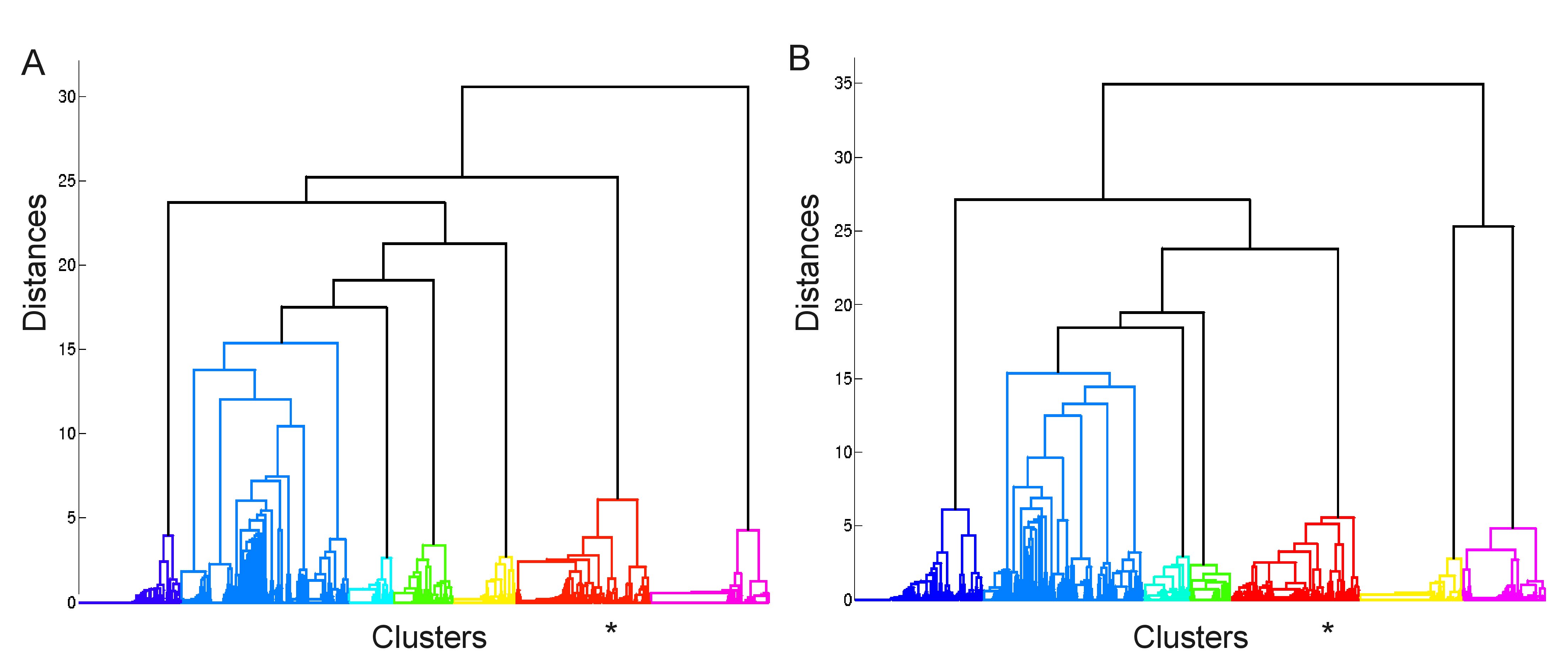} 
\caption{\textbf{Temporal changes in the cluster structure of the patent system.} Dendrograms representing the results of the hierarchical Ward clustering of patents in subcategory 11, based on their citation vector similarity on Jan. 1, 1994 (18833 patents in graph A) and Dec. 31, 1999 (25624 in graph B).  The x axis denotes a list of patents in subcategory 11, while the distances between them, as defined by the citation vector similarity, are drawn on the $y$ axis. (Patents separated by 0 distance form thin lines on the $x$ axis.) The 7 colors of the dendrogram correspond to the 7 most widely separated clusters. While the overall structure is  similar in 1994 and 1999, interesting structural changes emerged in this period. The cluster marked with the red color and asterix approximately corresponds to the new class 442, which was established in 1997, but was clearly identifiable by our clustering algorithm as early as 1991.
\label{fig:dendrogram}}
\end{centering}
\end{figure*}

Temporal changes in the cluster structure of the patent system can be detected in the changes of dendrograms. We present the dendrogram structure of the subcategory 11 at two different times (Figure \ref{fig:dendrogram}). Comparing the hierarchical structure in 1994 and 2000, we can observe both quantitative changes, when only the height of the branching point (branch separation distance) changed, and qualitative changes, when a new branching point has appeared.

In general, the hierarchical clustering is very sensitive and even the structure of large branches could be changed by changing only a few elements in the basic set. In spite of this general sensitivity, the main branches in the presented dendrograms were remarkably stable through the significant temporal changes during the development - they were easily identifiable from 1991 to 2000, during which time a significant increase took place in the underlying set, as can be seen from Table 1. The main branches and their large scale structure were also stable against minor modifications of the algorithm, such as the weighting method of the given citations during the citation vector calculation. These observations show, that the large branches are well identified real structures. However, this reliability does not necessarily hold for smaller cluster patterns, which are often the focus of interest. Thus, if a small cluster is identified as a candidate for becoming a new  field by our algorithm, it should be considered as a suggestion and should be evaluated by experts to verify the result.

really existing structures. Although, this does not necessarily hold for smaller cluster patterns, often in the focus of interest. Thus if a small cluster identified as a candidate becoming a new field by our algorithm, it should be considered as a suggestion and should be revised by experts to verify the result.

\subsection{The emergence of new classes: an illustration}

\begin{figure*}
\begin{centering}
\includegraphics[width=0.85\textwidth]{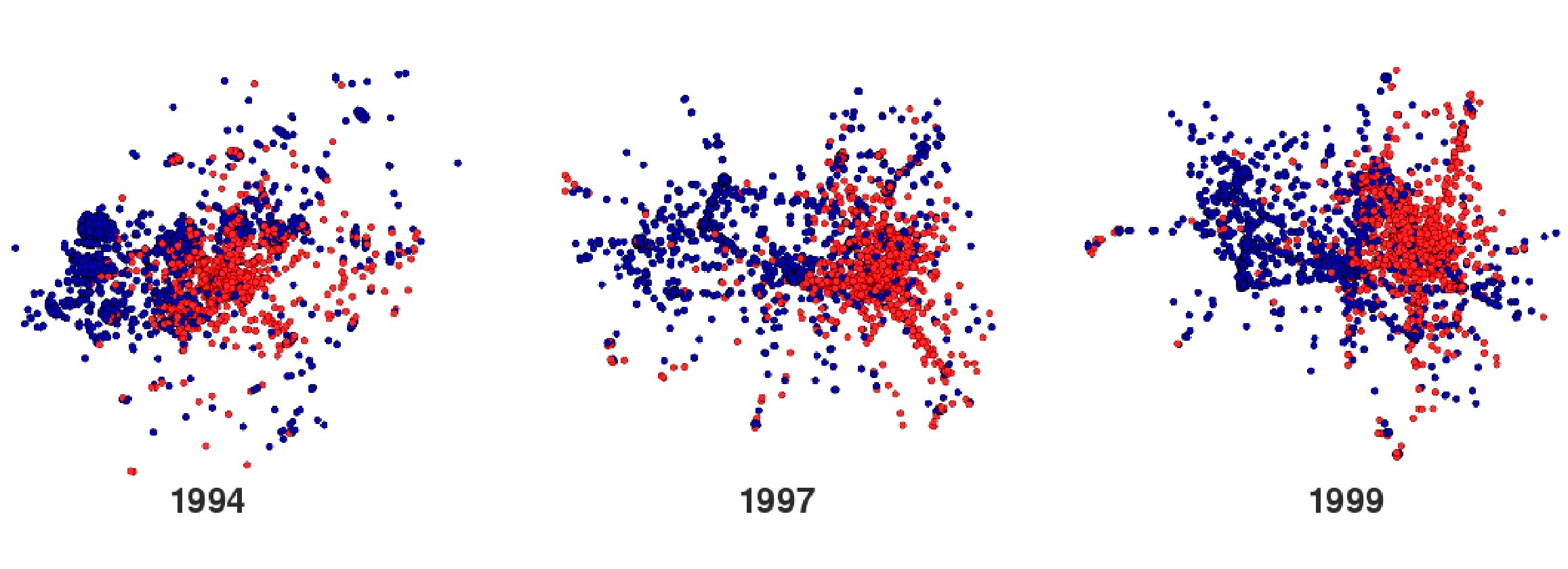}
\caption{\textbf{An example of the splitting process in the citation space, underlying the formation of a new class.} In the 2D projection of the 36 dimensional citation space, position of the circles denote the position of the patents in subcategory 11 in the citation space in three different stages of the separation process (Jan.\,1,1994, Jan.\,1,1997, Dec.\,31,1999). Red circles show those patents which were reclassified into the newly formed class 442, during the year 1997. The rest of the patents which preserved their classification after 1997 are denoted by blue circles. Precursors of the separation appear well before the official establishment of the new class. \label{fig:splitting}}
\end{centering}
\end{figure*}

The most important preliminary validation of our methodology is our ability to ``predict'' the emergence of a new technology class that was eventually identified by the USPTO.  As we mentioned earlier, the USPTO classification scheme not only provides the basis for the NBER subcategories that define our citation vector, it also provides a number of natural experiments to test the predictive power of our clustering method.  When the USPTO identifies a new technological category it defines a new class and then may reclassify earlier patents that are now recognized to have been part of that incipient new technological category. (Recall that there are many more USPTO classes than NBER subcategories -- within a given subcategory there are patents from a number of USPTO classes.) If our clustering method is sensitive to the emergence of new technological fields, we might hope that it will identify new technological branches before the USPTO recognizes their existence and defines new classes.

\begin{figure*}
\begin{centering}
\includegraphics[width=0.85\textwidth]{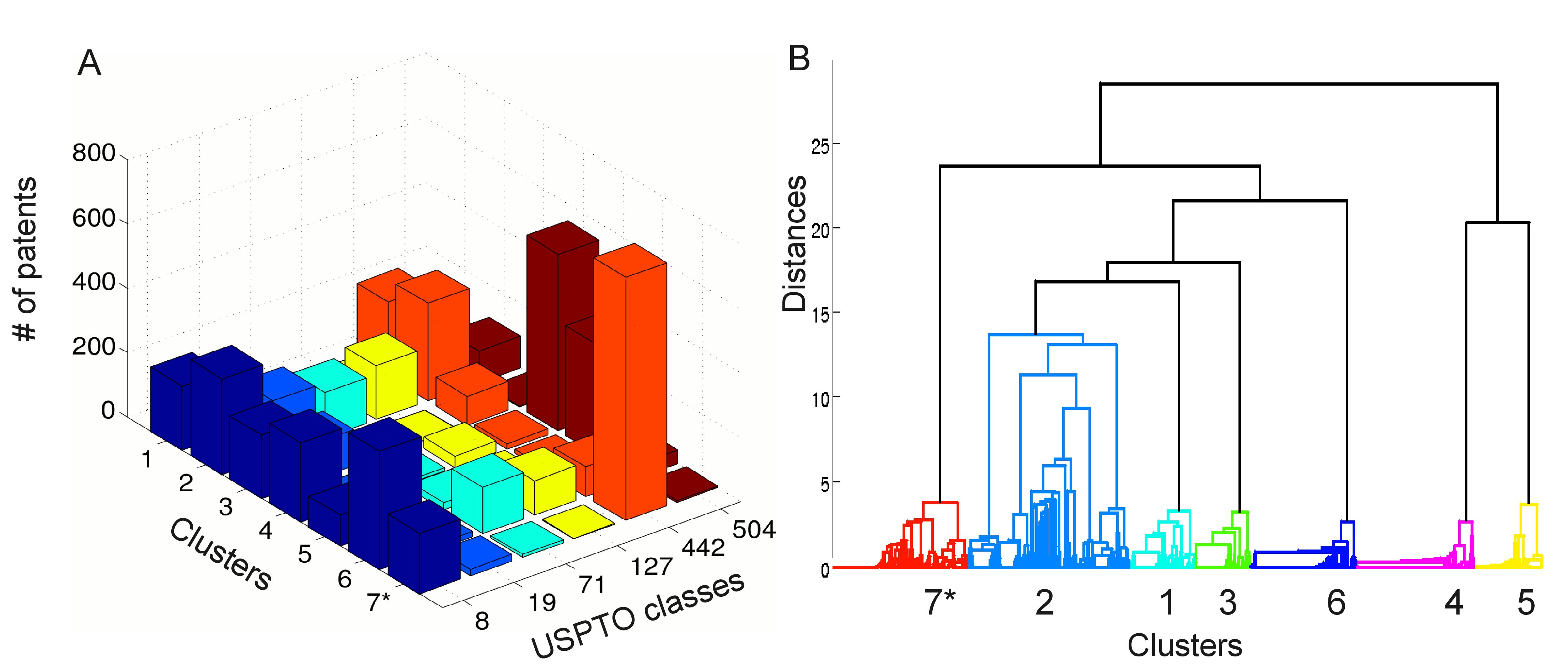}
\caption{\textbf{Separation of the patents by clustering in the citation space, based on the Jan.\,1,1991 data.} {\bf A}: Distribution of the patents issued before 1991 in the subcategory 11, within the 6 official classes in 1997 on the class axis (also marked with different colors) and within the 7 clusters in the citation space. The clustering algorithm collected the majority of those patents which were later reclassified into the newly formed class {\bf 442} (orange line) into the cluster {\bf 7} (marked with an asterisk). Vice-verse, the cluster {\bf 7} contains almost exclusively those patents which were later reclassified. Thus, we were able to identify the precursors of the emerging new class by clustering in the citation space. {\bf B}: The dendrogram belonging to the hierarchical clustering of the patents in the subcategory 11 in year 1991 shows that the branch which belongs to the cluster {\bf 7} is the most widely separated branch of the tree. The coloring here refers to the result of the clustering, unlike graph {\bf A} where coloring marks the USPTO classes. \label{fig:separation}}
\end{centering}
\end{figure*}

Figures \ref{fig:splitting} and \ref{fig:separation} illustrate the emergence of class 442, which was not defined by the USPTO until 1997.  Figure \ref{fig:splitting} shows how patents that will eventually be reclassified into class 442 can be seen to be splitting off from other patents in subcategory 11 as early as 1991. The visually recognizable cluster of patents in Figure \ref{fig:splitting} that will later be reclassified into class 442 can be identified by the Ward method with cutoff at 7 clusters in 1991, as is shown in Figure \ref{fig:separation}. The histogram in Figure \ref{fig:separation} shows the frequency of patents with a given cluster number and USPTO class.  Patents that will eventually be reclassified into class 442 are already concentrated in cluster 7. The Pearson-correlation between the class 442 and the correspondig clusters in our analysis resulted in high values: 0.9106 in 1991; 0.9005 in 1994; 0.8546 in 1997 and 0.9177 in the end of 1999. This example thus demonstrates that the citation vector can play the role of a predictor: emerging patent classes can be identified.

\section{Discussion}

Patent citation data seems to be a goldmine of new insights into the development of technologies, since  it  represents, even with noise, the innovation process. Scholars have long sought to understand technological change using evolutionary analogies, describing it as a process of recombination of already existing technologies \citep{schumpter39,usher54,henderson90,weitzman96,hargadon97}. Inventions are often described as combinations of prior technologies. ``...For example, one might think of the automobile as a combination of the bicycle, the horse carriage, and the internal combustion engine'' \citep{podolny95,podolny96,fleming01,fleming01b}. This feature of technological advance is well-recognized in patent law and has been the subject of recent Supreme Court attention, see KSR Int'l Co. v. Teleflex, Inc., 550 U.S. 398 (2007).  Our methodology exploits and tests this perspective by using the role a particular patented technology plays in combining existing technological fields to detect the emergence of new technology areas.  In this respect our method improves upon clustering methods based entirely on the existence of a citation between two articles or patents by incorporating locally-generated information (the patent category) relevant to the meaning of a given citation.  Here we present a proof of concept for the method.  In future work we will scan the patent citation network more broadly to identify clusters that may reflect the incipient development of new technological ``hot spots.''

We recognize that our method has a number of limitations. An important limitation is the time lag between the birth of a new technology and its appearance in the patent databases as reflected in the accumulation of citations. Cs\'ardi et al.\ \citep{csardi09} showed that the probability that an existing patent will be cited by a new patent peaks about 15 months after issuance. This time lag seems to show little variance across different fields. We may therefore expect that a fair amount of information about the use of a patented technology may have accumulated during that time.  It is also the case, however, that the citation probability exhibits a long tail, so that patents may continue to receive additional citations (potentially from different technological areas) over very long times.  Clearly the usefulness of the methodology as a predictor will depend on its ability to identify emerging technology areas before they are otherwise recognized.  In the specific example we explored here we were able to identify a new class well before its official recognition by the USPTO.  In future work, we will seek to determine the time difference between the detection of the first signs that a new cluster is emerging and the official formation of the new class for other cases to determine whether there is a characteristic time lag and, if so, whether it varies among technological categories.

The new method combines objective and subjective features.  The citations themselves (the links between citing and cited
documents) are based on similar technology/application concepts, and can be viewed as more or less objective quantities.  The citation vector bases are the manually assigned categories, and can be viewed as more subjective quantities.  Thus, in
some sense, the approach is a marriage of two types of taxonomies, each having many possible variations.  We suppose that USPTO classification catches {\bf sufficient} details to reflect in  advance some change in its structure.  If the resolution is low, i.e. only the larger categories are used we can identify only rare and really deep/fundamental changes.

Our methodology also oversimplifies the patent citation network in many ways. For example,  technological fields are not homogeneous with respect to number of patents, average number of citations per patent, and so forth.  Future work should explore the implications of these differences, especially the consequences of them to the weighting we applied when calculating the citation vector.

Finally, because there is no way to determine a priori the appropriate number of clusters for a given subset of the citation network, the method described provides only candidates for the identification of a new technological branch. To put it another way, we offer a decision support system: we are able to identify candidates for hot spots of technological development that are worthy of attention.

In future work, we also hope to use the method to examine, from a more theoretical perspective, the specific mechanisms of technological branching.  New technological branches can be generated either by a single (cluster dynamical) elementary event or by combinations of such events. For example, a new cluster might arise from a combination of merging and splitting. By examining historical examples, we hope to observe how the elementary events interact to build the recombination process and identify the typical ``microscopic mechanisms'' underlying new class formation.  This more ambitious research direction is grounded in a hypothesis that social systems are causal systems -- complex systems with circular causality and feedback loops \citep{erdi07,erdi10} -- whose statistical properties may allow us to uncover rules that govern their development (for similar attempts see Leskovec and coworkers (\citeyear{leskovec05}), and \citet{berlingerio09}). ``...Analogously to what happened in physics, we are finally in the position to move from the analysis of the ``social atoms'' or ``social molecules'' (i.e., small social groups) to the quantitative analysis of social aggregate states...'' \citep{vespignani09}.  Our study of the specific example of the patent citation network may thus help in the long run to advance our understanding of how complex social systems evolve.

\titlespacing{\section}{0pt}{12pt}{5pt}

\section*{Acknowledgments}
PE thanks  the Henry Luce Foundation and the Toyota Research Institue for their support. KJS acknowledges the
generous support of The Filomen D'Agostino and Max E. Greenberg Research
Fund. Thanks for F\"ul\"op Bazs\'o, Mih\'aly B\'anyai, Judit Szente,
Bal\'azs Ujfalussy for discussions.

The final publication is available at www.springerlink.com.\\
\sloppy
\url{http://www.springerlink.com/openurl.asp?genre=article&id=doi:10.1007/s11192-012-0796-4} 


\bibliographystyle{spbasic}
\bibliography{patent1_corr}

\begin{thebibliography}{85}
\providecommand{\natexlab}[1]{#1}
\providecommand{\url}[1]{{#1}}
\providecommand{\urlprefix}{URL }
\expandafter\ifx\csname urlstyle\endcsname\relax
  \providecommand{\doi}[1]{DOI~\discretionary{}{}{}#1}\else
  \providecommand{\doi}{DOI~\discretionary{}{}{}\begingroup
  \urlstyle{rm}\Url}\fi
\providecommand{\eprint}[2][]{\url{#2}}

\bibitem[{Alcacer and Gittelman(2006)}]{alcacer06}
Alcacer J, Gittelman M (2006) How do i know what you know? patent examiners and
  the generation of patent citations. Review of Economics and Statistics
  88(4):774--779

\bibitem[{Almeida and Kogut(1997)}]{almeida-kogut97}
Almeida P, Kogut B (1997) The exploration of technological diversity and the
  geographic localization of innovation. Small Business Economics 9:21--31

\bibitem[{Berlingerio et~al(2009)Berlingerio, Bonchi, Bringmann, and
  Gionis}]{berlingerio09}
Berlingerio M, Bonchi F, Bringmann B, Gionis A (2009) Mining graph evolution
  rules. In: Buntine W, Grobelnik M, Mladenic D, Shawe-Taylor J (eds) Machine
  Learning and Knowledge Discovery in Databases, Springer, European Conference
  on Machine Learning and Knowledge Discovery in Databases, pp 115--130

\bibitem[{Blondel et~al(2008)Blondel, Guillaume, Lambiotte, and
  Lefebvre}]{blondeletla08}
Blondel V, Guillaume JL, Lambiotte R, Lefebvre E (2008) Fast unfolding of
  communities in large networks. Journal of Statistical Mechanics: Theory and
  Experiment p P10008

\bibitem[{Breitzman(2007)}]{breitzman07}
Breitzman A (2007) The emerging clusters project. Final Report, 1790
  {Analytics}, www.ntis.gov/pdf/Report-EmergingClusters.pdf

\bibitem[{Chang et~al(2009)Chang, Lai, and Chang}]{chang09}
Chang S, Lai K, Chang S (2009) Exploring technology diffusion and
  classification of business methods: using the patent citation network.
  Technological Forecasting and Social Change 76(1):107--117

\bibitem[{Chen et~al(2010)Chen, Ibekwe-SanJuan, and Hou}]{chenetal10}
Chen C, Ibekwe-SanJuan F, Hou J (2010) The structure and dynamics of
  co-citation clusters: A multiple-perspective co-citation analysis. Journal of
  the American Society for Information Science and Technology 61:1386--1409

\bibitem[{Criscuolo and B.Verspagen(2008)}]{criscuolo08}
Criscuolo P, BVerspagen (2008) Does it matter where patent citations come from?
  inventor versus examiner citations in european patents. Research Policy
  37:1892--1908

\bibitem[{Cs\'ardi et~al(2009)Cs\'ardi, Strandburg, Tobochnik, and
  \'Erdi}]{csardi09}
Cs\'ardi G, Strandburg K, Tobochnik J, \'Erdi P (2009) Chapter 10. the inverse
  problem of evolving networks -- with application to social nets. In:
  Bollob\'as B, Kozma R, Mikl\'os D (eds) Handbook of Large-Scale Random
  Networks, Springer-Verlag, Berlin, Heidelberg, New York, pp 409--443

\bibitem[{Day and Schoemaker(2005)}]{day-smaker05}
Day G, Schoemaker P (2005) Scanning the periphery. Harvard Business Review pp
  1--12

\bibitem[{Debackere et~al(2002)Debackere, Verbeek, Luwel, and
  Zimmermann}]{debackere02}
Debackere K, Verbeek A, Luwel M, Zimmermann E (2002) The multiple uses of
  technometric indicators. International Journal of Management Reviews
  4:213--231

\bibitem[{van Dongen(2000)}]{dongen00}
van Dongen S (2000) A cluster algorithm for graphs. Technical Report INS-R0010,
  National Research Institute for Mathematics and Computer Science in the
  Netherlands, Amsterdam

\bibitem[{Duguet and MacGarvie(2005)}]{duguet05}
Duguet E, MacGarvie M (2005) How well do patent citations measure flows of
  technology? {E}vidence from {F}rench innovation surveys. Economics of
  Innovation and New Technology 14(5):375--393

\bibitem[{\'Erdi(2007)}]{erdi07}
\'Erdi P (2007) Complexity Explained. Springer-Verlag, Berlin, Heidelberg

\bibitem[{\'Erdi(2010)}]{erdi10}
\'Erdi P (2010) Scope and limits of predictions by social dynamic models:
  Crisis, innovation, decision making. Evolutionary and Institutional Economic
  Review 7:21--42

\bibitem[{Fleming(2001)}]{fleming01}
Fleming L (2001) Recombinant uncertainty in technological search. Management
  Science 47(1):117--132

\bibitem[{Fleming(2004)}]{fleming04}
Fleming L (2004) Science as a map in technological search. Strategic Management
  Journal 25:909--928

\bibitem[{Fleming and Sorenson(2001)}]{fleming01b}
Fleming L, Sorenson O (2001) Technology as a complex adaptive system: evidence
  from patent data. Research Policy 30:1019--1039

\bibitem[{Fleming et~al(2006)Fleming, Juda, and III}]{fleming06}
Fleming L, Juda A, III CK (2006) Small worlds and regional innovation. Harvard
  Business School Working Paper Series, No. 04--008, available at
  http://ssrn.com/abstract=892871

\bibitem[{Fontana et~al(2009)Fontana, Nuvolari, and Verspagen}]{fontana09}
Fontana R, Nuvolari A, Verspagen M (2009) Mapping technological trajectories as
  patent citation networks. an application to data communication standards.
  Economics of Innovation and New Technology 18:311--336

\bibitem[{Garfield(1983)}]{garfield83}
Garfield E (1983) Citation Indexing -- Its Theory and Application in Science,
  Technology and Humanities. ISI Press, Philadelphia

\bibitem[{Garfield(1993)}]{garfield93}
Garfield E (1993) Co-citation analysis of the scientific literature: Henry
  small on mapping the collective mind of science. Current Contents 19:3--13

\bibitem[{Girvan and Newman(2002)}]{girvan02}
Girvan M, Newman M (2002) Community structure in social and biological
  networks. PNAS 99(12):7821--7826

\bibitem[{Hagedoorn and Cloodt(2003)}]{hagedoorn03}
Hagedoorn J, Cloodt M (2003) Measuring innovative performance: Is there an
  advantage in using multiple indicators? Research Policy 32:1365--1379

\bibitem[{Hall et~al(2001)Hall, Jaffe, and Trajtenberg}]{hall01}
Hall B, Jaffe A, Trajtenberg M (2001) The {NBER} patent citation data file:
  Lessons, insights and methodological tools. Working Paper 8498, National
  Bureau of Economic Research

\bibitem[{Hargadon and Sutton(1997)}]{hargadon97}
Hargadon A, Sutton R (1997) Technology brokering and innovation in a product
  development firm. Administrative Science Quarterly 42:716--749

\bibitem[{Harhoff et~al(1999)Harhoff, Narin, Scherer, and Vopel}]{harhoff99}
Harhoff D, Narin F, Scherer F, Vopel K (1999) Citation frequency and the value
  of patented inventions. The Review of Economics and Statistics 81:511--515

\bibitem[{Henderson and Clark(1990)}]{henderson90}
Henderson R, Clark K (1990) Architectural innovation: The reconfiguration of
  existing product technologies and the failure of established firms.
  Administrative Science Quarterly 35(1):9--30

\bibitem[{Huang et~al(2003)Huang, Chen, Yip, Ng, Guo, Chen, and Roco}]{huang03}
Huang Z, Chen H, Yip A, Ng G, Guo F, Chen ZK, Roco M (2003) Longitudinal patent
  analysis for nanoscale science and engineering: Country, institution and
  technology field. Journal of Nanoparticle Research 5:333--363

\bibitem[{Huang et~al(2004)Huang, Chen, Chen, and Roco}]{huangetal04}
Huang Z, Chen H, Chen ZK, Roco M (2004) International nanotechnology
  development in 2003: Country, institution, and technology field analysis
  based on uspto patent database. Journal of Nanoparticle Research 6:325--354

\bibitem[{Jaffe and Trajtenberg(2005)}]{jaffe-trajt05}
Jaffe A, Trajtenberg M (2005) Patents, Citations \& Innovations -- a Window on
  the Knowledge Economy. MIT Press, Cambridge

\bibitem[{Kajikawa and Takeda(2008)}]{kajikawa08c}
Kajikawa Y, Takeda Y (2008) Structure of research on biomass and bio-fuels: A
  citation-based approach. Technological Forecasting and Social Change
  75:1349--1359

\bibitem[{Kajikawa et~al(2008)Kajikawa, Usui, Hakata, Yasunaga, and
  Matsushima}]{kajikawa08a}
Kajikawa Y, Usui O, Hakata K, Yasunaga Y, Matsushima K (2008) Sructure of
  knowledge in the science and technology roadmaps. Technological Forecasting
  and Social Change 75:1--11

\bibitem[{Kostoff and Geisler(2007)}]{kostoff-geisler07}
Kostoff R, Geisler E (2007) The unintended consequences of metrics in
  technology evaluation. Journal of Infometrics 1:103--114

\bibitem[{Kostoff and Schaller(2001)}]{kostoff-schaller01}
Kostoff R, Schaller R (2001) Science and technology roadmaps. IEEE Transactions
  on Engineering Management 48:132--143

\bibitem[{Kostoff et~al(2006)Kostoff, Stump, Johnson, Murday, Lau, and
  Tolles}]{kostoff06}
Kostoff R, Stump J, Johnson D, Murday J, Lau C, Tolles W (2006) The structure
  and infrastructure of the global nanotechnology literature. Journal of
  Nanoparticle Research 8:301--321

\bibitem[{Lai and Wu(2005)}]{laiwu05}
Lai K, Wu SJ (2005) Using the patent co-citation approach to establish a new
  patent classification system. Information Processing and Management
  41:313--330

\bibitem[{Lanjouw and Schankerman(2004)}]{lanjouw-schankerman04}
Lanjouw J, Schankerman M (2004) Patent quality and research productivity:
  Measuring innovation with multiple indicators. The Economic Journal
  114(495):441--465

\bibitem[{Lee et~al(2010)Lee, Su, and Wu}]{lee10}
Lee PC, Su HN, Wu FS (2010) Quantitative mapping of patented technology -- the
  case of electrical conducting polymer nanocomposite. Technological
  Forecasting and Social Change 77(3):466--478

\bibitem[{Leskovec et~al(2005)Leskovec, Kleinberg, and Faloutsos}]{leskovec05}
Leskovec J, Kleinberg J, Faloutsos C (2005) Graphs over time: densification
  laws, shrinking diameters and possible explanations. In: {KDD} 2005:
  Proceedings of the eleventh {ACM SIGKDD} international conference on
  {Knowledge} discovery in data mining, ACM, New York, NY, pp 177--187

\bibitem[{Leydesdorff(2008)}]{leydesdorff08}
Leydesdorff L (2008) Patent classifications as indicators of intellectual
  organization. Journal of the American Society for Information Science and
  Technology 59(10):1582--1597

\bibitem[{McMillanm et~al(2000)McMillanm, Narin, and Deeds}]{mcmillan00}
McMillanm G, Narin F, Deeds D (2000) An analysis of the critical role of public
  science in innovation: the case of biotechnology. Research Policy 29:1--8

\bibitem[{Meyer(2000)}]{meyer00}
Meyer M (2000) What is special about patent citations? differences between
  scientific and patent citations. Scientometrics 49:93--123

\bibitem[{Meyer(2001)}]{meyer01}
Meyer M (2001) Patent citation analysis in a novel field of technology: An
  exploration of nano-science and nano-technology. Scientometrics 51:163--183

\bibitem[{Milman(1994)}]{milman94}
Milman B (1994) Individual cocitation clusters as nuclei of complete and
  dynamic infrometric models of scientific and technological areas.
  Scientometrics 31:45--57

\bibitem[{Moed(2005)}]{moed05}
Moed H (2005) Citation Analysis in Research Evaluation. Springer, Dordrecht,
  Netherlands

\bibitem[{Mogee and Kolar(1998{\natexlab{a}})}]{mogee-kolar98b}
Mogee M, Kolar R (1998{\natexlab{a}}) Patent citation analysis of allergan
  pharmaceutical patents. Expert Opinion on Therapeutic Patents
  8(10):1323--1346

\bibitem[{Mogee and Kolar(1998{\natexlab{b}})}]{mogee-kolar98a}
Mogee M, Kolar R (1998{\natexlab{b}}) Patent citation analysis of new chemical
  entities claimed as pharmaceuticals. Expert Opinion on Therapeutic Patents
  8(3):213--222

\bibitem[{Mogee and Kolar(1999)}]{mogee-kolar99}
Mogee M, Kolar R (1999) Patent co-citation analysis of eli lilly \& co.
  patents. Expert Opinion on Therapeutic Patents 9(3):291--305

\bibitem[{Murray(2002)}]{murray02}
Murray F (2002) Innovation as co-evolution of scientific and technological
  networks: exploring tissue engineering. Research Policy 31:1389--1403

\bibitem[{Narin(1994)}]{narin94}
Narin F (1994) Patent bibliometrics. Scientometrics 30:147--155

\bibitem[{Newman(2002)}]{newman02}
Newman M (2002) Assortative mixing in networks. Physical Review Letters
  89(20):208,701

\bibitem[{Newman(2006)}]{newman06}
Newman M (2006) Finding community structure in networks using the eigenvectors
  of matrices. Physical Review E 74:036,104

\bibitem[{Newman and Girvan(2004)}]{newman04}
Newman M, Girvan M (2004) Finding and evaluating community structure in
  networks. Physical Review E 69(026113)

\bibitem[{OuYang and Weng(2011)}]{ouyang11}
OuYang K, Weng C (2011) A new comprehensive patent analysis approach for new
  product design in mechanical engineering. Technological Forecasting and
  Social Change 78(7):1183--1199

\bibitem[{Palla et~al(2007)Palla, Barab\'asi, and Vicsek}]{palla07}
Palla G, Barab\'asi AL, Vicsek T (2007) Quantifying social group evolution.
  Nature 446:664--667

\bibitem[{Podolny and Stuart(1995)}]{podolny95}
Podolny J, Stuart T (1995) A role-based ecology of technological change. The
  American Journal of Sociology 100(5):1224--1260

\bibitem[{Podolny et~al(1996)Podolny, Stuart, and Hannan}]{podolny96}
Podolny J, Stuart T, Hannan M (1996) Networks, knowledge, and niches:
  Competition in the worldwide semiconductor industry, 1984-1991. The American
  Journal of Sociology 102(3):659--689

\bibitem[{Pons and Latapy(2006)}]{pons06}
Pons P, Latapy M (2006) Computing communities in large networks using random
  walks. Journal of Graph Algorithms and Applications 10:191--218

\bibitem[{Pyka and Scharnhost(2009)}]{pyka09}
Pyka A, Scharnhost A (2009) Innovation Networks. New Approaches in Modelling
  and Analyzing. Springer-Verlag, Berlin, Heidelberg

\bibitem[{Sampat(2004)}]{sampat04}
Sampat B (2004) Examining patent examination: An analysis of examiner and
  application generated prior art. Working Paper, Prepared for NBER Summer
  Institute

\bibitem[{Sampat and Ziedonis(2002)}]{sampat-ziedonis02}
Sampat B, Ziedonis A (2002) Cite seeing: Patent citations and the economic
  value of patents. Unpublished manuscript, from the author

\bibitem[{Saviotti(2005)}]{saviotti05}
Saviotti P (2005) On the co-evolution of technologies and institutions. In:
  Weber M, Hemmelskamp J (eds) Towards Environmental Innovations Systems,
  Springer, Berlin, Hidelberg

\bibitem[{Saviotti et~al(2003)Saviotti, de~Looze, and Maopertuis}]{saviotti03}
Saviotti P, de~Looze M, Maopertuis M (2003) Knowledge dynamics and the mergers
  of firms in the biotechnology based sectors. International Journal of
  Biotechnology 5(3--4):371--401

\bibitem[{Saviotti et~al(2005)Saviotti, de~Looze, and Maopertuis}]{saviotti05b}
Saviotti P, de~Looze M, Maopertuis M (2005) Knowledge dynamics, firm strategy,
  mergers and acquisitions in the biotechnology based sectors. Economics of
  Innovation and New Technology 14(1--2):103--124

\bibitem[{Schumpeter(1939)}]{schumpter39}
Schumpeter J (1939) Business Cycles. McGraw-Hill, New York

\bibitem[{Shibata et~al(2008)Shibata, Kajikawa, Takeda, and
  Matsushima}]{shibata08}
Shibata N, Kajikawa Y, Takeda Y, Matsushima K (2008) Detecting emerging
  research fronts based on topological measures in citation networks of
  scientific publications. Technovation 28:758--775

\bibitem[{Shibata et~al(2010)Shibata, Kajikawa, and Sakata}]{shibata10}
Shibata N, Kajikawa Y, Sakata I (2010) Extracting the commercialization gap
  between science and technology -- case study of a solar cell. Technological
  Forecasting and Social Change 77:1147--1155

\bibitem[{Shibata et~al(2011)Shibata, Kajikawa, Takeda, Sakata, and
  Matsushima}]{shibata11}
Shibata N, Kajikawa Y, Takeda Y, Sakata I, Matsushima K (2011) Detecting
  emerging research fronts in regenerative medicine by the citation network
  analysis of scientific publications. Technological Forecasting and Social
  Change 78:274--282

\bibitem[{Small(1973)}]{small73}
Small H (1973) Cocitation in scientific literature: New measure of relationship
  between two documents. Journal of The American Society For Information
  Science 24:265--269

\bibitem[{Small(2006)}]{small06}
Small H (2006) Tracking and predicting growth areas in science. Scientometrics
  68:595--610

\bibitem[{Sood and Tellis(2005)}]{sood-tellis05}
Sood A, Tellis G (2005) Technological evolution and radical innovation. Journal
  of Marketing 69:152--168

\bibitem[{Sorenson et~al(2006)Sorenson, Rivkin, and Fleming}]{sorenson06}
Sorenson O, Rivkin J, Fleming L (2006) Complexity, networks and knowledge flow.
  Research Policy 35(7):994--1017

\bibitem[{Sternitzke(2009)}]{sternitzke09}
Sternitzke C (2009) Patents and publications as sources of novel and inventive
  knowledge. Scientometrics 79:551--561

\bibitem[{Strandburg et~al(2007)Strandburg, Cs\'ardi, Tobochnik, \'Erdi, and
  Zal\'anyi}]{strandburg07}
Strandburg K, Cs\'ardi G, Tobochnik J, \'Erdi P, Zal\'anyi L (2007) Law and the
  science of networks: An overview and an application to the ``patent
  explosion''. Berkeley Technology Law Journal 21:1293

\bibitem[{Strandburg et~al(2009)Strandburg, Cs\'ardi, Tobochnik, \'Erdi, and
  Zal\'anyi}]{strandburg09}
Strandburg K, Cs\'ardi G, Tobochnik J, \'Erdi P, Zal\'anyi L (2009) Patent
  citation networks revisited: signs of a twenty-first century change? North
  Carolina Law Review 87:1657--1698

\bibitem[{Strumsky et~al(2005)Strumsky, Lobo, and Fleming}]{strumsky05}
Strumsky D, Lobo J, Fleming L (2005) Metropolitan patenting, inventor
  agglomeration and social networks: A tale of two effects. SFI Working Paper
  No. 05-02-004, available at
  http://www.santafe.edu/media/workingpapers/05-02-004.pdf

\bibitem[{Tijssen(2001)}]{tijssen01}
Tijssen R (2001) Global and domestic utilization of industrial relevant
  science: patent citation analysis of science-technology interactions and
  knowledge flows. Research Policy 30:35--54

\bibitem[{Usher(1954)}]{usher54}
Usher A (1954) A History of Mechanical Invention. Dover, Cambridge

\bibitem[{Verbeek et~al(2002)Verbeek, Debackere, Luwel, and
  Zimmermann}]{verbeek02}
Verbeek A, Debackere K, Luwel M, Zimmermann E (2002) Measuring progress and
  evolution in science and technology -- {I}: The multiple uses of bibliometric
  indicators. International Journal of Management Reviews 4(2):179--211

\bibitem[{Vespignani(2009)}]{vespignani09}
Vespignani A (2009) Predicting the behavior of techno-social systems. Science
  325(5939):425--428

\bibitem[{Wallace et~al(2009)Wallace, Gingras, and Duhon}]{wallaceetal09}
Wallace M, Gingras Y, Duhon R (2009) A new approach for detecting scientific
  specialties from raw cocitation networks. Journal of the American Society for
  Information Science and Technology 60(2):240--246

\bibitem[{Ward(1963)}]{ward63}
Ward J (1963) Hierarchical grouping to optimize an objective function. Journal
  of the American Statistical Association 58(301):236--244

\bibitem[{Weitzman(1996)}]{weitzman96}
Weitzman M (1996) Hybridizing growth theory. American Economic Review
  86(2):207--12

\bibitem[{Weng et~al(2010)Weng, Chen, Hsu, and Chien}]{weng10}
Weng C, Chen WY, Hsu HY, Chien SH (2010) To study the technological network by
  structural equivalence. Journal of High Technology Management Research
  21:52--63

\end{thebibliography}

\section{Biographies}

       P\'eter \'Erdi is the is the Henry R. Luce Professor of Complex
Systems studies in Kalamazoo College,  and also a research professor at the
Wigner Research Centre for Physics, Hungarian Academy of Sciences
He has been working on the fields of computational neuroscience and computational social
sciences.

Kinga Makovi is a PhD student in the Department of Sociology at Columbia
University, and holds an MS in mathematical economics from Corvinus
University in Budapest (2010). Her interests
include social networks, quantitative methods and simulation techniques in
social sciences.

Zolt\'an Somogyv\'ari is a senior research fellow at the Wigner Research Centre for Physics, Hungarian Academy of Sciences. He is an expert in developing new methods of analysing large data sets.

Katherine Strandburg  is a Professor of Law at New York University School
of Law. Her teaching and research activities are in the areas of
intellectual property law, cyberlaw, and information privacy law. Prior to
her legal caireer, she was a research physicist at Argonne National
Laboratory, having received her Ph.D. in Physics  from Cornell University.

Jan Tobochnik is the Dow Distinguished Professor of Natural Science in
Kalamazoo College. He also served as the Editor of the {\it American Journal of
Physics}. His research involves using computer simulations to understand a
wide variety of systems. In the last decade he has been involved in
investigating the structural properties of social networks.

P\'eter Volf got his master degree in the Department of Measurement and
Information Systems of the Budapest University of Technology and
Economics. His main interest is now developing efficient clustering
algorithms. He is now with Nokia Siemens Networks, Budapest.

L\'aszl\'o Zal\'anyi is a senior research fellow at the Wigner Research Centre for Physics, Hungarian Academy of Sciences. He is an expert in developing new methods of analysing large data sets. His research areas are the application of
stochastic methods to neural and social systems, and network theory.

\end{document}